\documentclass{article}

\PassOptionsToPackage{numbers,sort&compress}{natbib}

\usepackage[main, final]{neurips_2026}
\usepackage{graphicx}
\usepackage{amsmath,amssymb}
\usepackage{multirow}
\usepackage{array}
\usepackage{subcaption}
\usepackage{caption}

\usepackage[utf8]{inputenc} 
\usepackage[T1]{fontenc}    
\usepackage{hyperref}       
\usepackage{url}            
\usepackage{booktabs}       
\usepackage{amsfonts}       
\usepackage{nicefrac}       
\usepackage{microtype}      
\usepackage{xcolor}         

\title{DELTA-TTS: Adapting Autoregressive Model into \\ Diffusion Language Model for Text-to-Speech}   
\author{%
  Junwon Moon\textsuperscript{1}, Yejin Lee\textsuperscript{1}, Seungbeom Kim\textsuperscript{1}, Hoseong Ahn\textsuperscript{1}, Sewoong Park\textsuperscript{1},\\
  \textbf{Heeseung Kim\textsuperscript{2}, Kyuhong Shim\textsuperscript{1}} \\
  \textsuperscript{1}Sungkyunkwan University 
  \textsuperscript{2}University of Seoul \\
  \texttt{\{mppn98, khshim\}@skku.edu} \\
}

\begin{document}

\maketitle

\begin{abstract}
Autoregressive (AR) text-to-speech (TTS) models generate discrete speech tokens sequentially, which makes inference slow and can degrade robustness, since local errors propagate to later positions and can escalate into hallucination.
This limitation stems from their left-to-right AR commitment: each token must be determined before future speech-token context is available.
However, such ordering is not an inherent requirement for TTS, since the model receives the full input text before synthesis.
In this paper, we introduce DELTA-TTS, a lightweight LoRA-based adaptation framework that converts a pretrained AR TTS model into a discrete diffusion language model (dLLM) for confidence-ordered speech-token decoding.
To better capture the local structure of speech, DELTA-TTS incorporates a convolution module that injects local acoustic context, together with a $1/t$-weighted training objective and a time-shifted inference schedule that together defer low-confidence positions to later steps.
Trained on only 585 hours of LibriTTS, DELTA-TTS achieves a 1.75\% WER on Seed-TTS test-en, outperforming its AR backbone while generating tokens 3.3 times faster.
Further analysis shows that DELTA-TTS produces sharper text--speech alignment, increases overall decoding confidence, and mitigates the hallucinations observed in AR generation.
Audio samples, inference code, and model weights are available at \url{https://slpr-lab.github.io/DELTA-TTS}.

\end{abstract}

\section{Introduction}\label{sec:intro}

Recent autoregressive (AR) text-to-speech (TTS) models synthesize highly natural speech using large language model (LLM) backbones that predict discrete speech tokens~\cite{cosyvoice3, cosyvoice2, hu2026qwen3, llasa}.
Despite their strong performance, AR TTS models have two fundamental limitations.
First, token-by-token generation leads to slow inference. 
This cost is particularly pronounced for speech, which requires far more tokens than text to represent the same content.
Second, unidirectional generation conditions each token only on past context.
Because neighboring speech tokens strongly depend on one another, the AR model's inability to attend to future context becomes a structural weakness.

\begin{figure*}[t]
  \centering
  \includegraphics[width=1.0\linewidth]{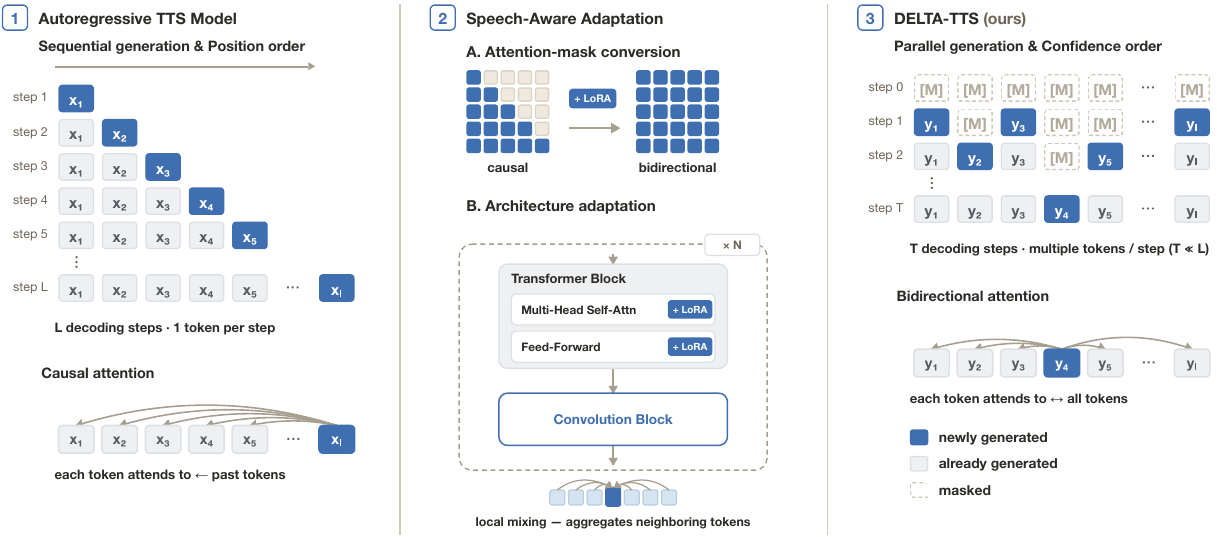}
  \caption{Overview of DELTA-TTS. A pretrained AR TTS backbone is converted into a discrete diffusion language model with two adaptations: bidirectional attention and block-wise LoRA with a Conformer-style convolution module.}
  \label{fig:conversion}
  \vspace{0.2cm}
\end{figure*}

Formally, AR TTS models the conditional distribution $P(A \mid T) = P(a_1, a_2, \dots \mid T)$, where $A$ is a speech-token sequence and $T$ is the input text.
The AR formulation factorizes this distribution as $P(a_1 \mid T)\,P(a_2 \mid a_1, T)\cdots$, forcing a fixed left-to-right generation order.
However, because the entire text $T$ is given in advance, this factorization is not the only option for producing $A$.
Since the learned model can only approximate the target conditional distribution, the decoding order affects how errors accumulate.
We hypothesize that decoding tokens in order of confidence, rather than in temporal (i.e., left-to-right) order, leads to more stable generation.
Interestingly, this hypothesis aligns closely with the generation process of discrete diffusion language models (dLLMs), which have recently emerged as a promising alternative to AR generation in the text domain~\cite{llada, mdlm, dream, shi2024simplified}.
In essence, dLLMs generate high-confidence tokens first and progressively complete the sequence through iterative refinement.
In this paper, we propose \textbf{DELTA-TTS}, a framework that adapts a pretrained AR TTS model into a dLLM, generating speech non-autoregressively while preserving the synthesis quality of the backbone.
The conversion process employs parameter-efficient LoRA-based modifications.
To better reflect the local and bidirectional structure of speech, we introduce two additional refinements.
First, we incorporate a Conformer-style convolution module to model the correlations among adjacent speech tokens.
Second, we combine a $1/t$-weighted training objective with a time-shifted inference schedule so that tokens with insufficient context can be deferred to later decoding steps.
With these refinements, the model learns a sharper text--speech alignment and reduces error accumulation caused by committing tokens too early.

We validate DELTA-TTS through both empirical evaluation and analysis.
Trained on only 585 hours of LibriTTS, DELTA-TTS achieves 1.75\% WER on Seed-TTS \emph{test-en} while generating tokens 3.3$\times$ faster, and our analysis shows how each refinement contributes to these gains.

Our contributions are as follows:
\begin{itemize}

\item We introduce DELTA-TTS, the first method to extend AR-to-dLLM conversion to speech. DELTA-TTS converts a pretrained AR TTS backbone into a non-autoregressive model through a lightweight LoRA-based modification that requires only $\sim$15\% additional parameters.

\item We propose two speech-aware designs for generation in confidence order rather than temporal order: a convolution module for locality, and a $1/t$-weighted training objective with a time-shifted schedule for deferring uncertain tokens to later steps.

\item We analyze the conversion from two perspectives. First, naive text-domain conversion substantially degrades performance, and the convolution module recovers it by sharpening the text--speech alignment.
Second, confidence-ordered decoding maintains high confidence across the sequence, mitigating hallucination.

\end{itemize}
\section{Related Work}\label{sec:related}

\subsection{Autoregressive and Non-Autoregressive TTS}
Modern zero-shot AR TTS systems formulate speech synthesis as language modeling over discrete tokens and are trained on large speech corpora~\cite{seedtts, cosyvoice3, valle}.
A growing line of work initializes their backbones from pretrained text language models~\cite{cosyvoice2, sparktts, fireredtts2}.
On top of these backbones, they attach speech embedding and output projection layers to process discrete speech tokens.
However, this paradigm inherits two limitations from AR language modeling.
First, generating an $N$-token speech sequence demands $N$ sequential forward passes, so inference latency grows linearly with the output length.
Second, attending only to past context prevents the model from exploiting future tokens, despite the strong local dependencies among neighboring speech tokens.

NAR TTS models instead generate all tokens within a fixed number of refinement steps, independent of the output length.
This parallel formulation is also naturally compatible with bidirectional attention, so the model can capture full-sequence context.
These methods can be broadly categorized into continuous-space approaches~\cite{f5tts, e2tts, le2023voicebox, lee2025dittotts, kim2026raonopentts} and discrete-token approaches~\cite{borsos2024soundstorm, ju2024naturalspeech, maskgct, wang2025metis}.
However, as with AR systems, NAR TTS models are typically trained from scratch on large-scale speech corpora, which incurs substantial computational cost.

To bridge the gap between these two paradigms, hybrid approaches combining AR and NAR generation have also been explored.
For example, DiTAR~\cite{ditar} performs autoregressive generation at the chunk level while employing diffusion-based decoding within each chunk.
While bidirectional attention operates within each chunk, generation across chunks remains sequential.
As a result, the latency bottleneck is only partially resolved.

\subsection{Discrete Diffusion Language Models}
Diffusion-based language modeling has emerged as a competitive NAR paradigm.
Early continuous-diffusion approaches~\cite{dieleman2022continuous, gong2023diffuseq, han2023ssd, diffusionlm} suffer from scalability issues and information loss when mapping continuous representations back to discrete tokens.
Subsequent work has therefore shifted toward discrete diffusion models that operate directly in the token space~\cite{d3pm, sedd, mdlm}.

Specifically, these models learn to predict masked tokens in a partially masked sequence, reversing a forward masking process.
At inference time, starting from a fully masked sequence, the model iteratively unmasks tokens in parallel and commits the most confident predictions at each step.
Building on this paradigm, recent dLLMs~\cite{arriola2025block, llada, dream, zhou2026dllm} have begun to narrow the quality gap with AR LLMs while retaining the efficiency advantages of parallel generation.

\subsection{AR-to-dLLM Conversion}
A recent line of research has shown that pretrained AR language models can be converted into dLLMs~\cite{diffullama, tess2, fastdllmv2, dream}.
These methods adapt pretrained AR weights through continual pretraining with a masked diffusion objective~\cite{mdlm}. 
Many of these methods further incorporate a shift operation that preserves the next-token prediction behavior of the AR model, so the pretrained weights transfer directly to parallel generation.
Because this conversion route avoids training from scratch, it substantially improves training efficiency. 
However, existing approaches have been developed primarily for text language models, and extending AR-to-dLLM conversion to TTS models remains unexplored.

\section{Method}\label{sec:method}

\subsection{Overview}\label{sec:overview}

DELTA-TTS adapts a pretrained AR TTS model into a dLLM, while keeping the speech tokenizer and the flow-matching decoder frozen so that the existing zero-shot inference pipeline can be reused without modification.
We use CosyVoice3~\cite{cosyvoice3} as the AR backbone, in which a Qwen2.5-0.5B language model produces a 25\,Hz semantic speech-token sequence and a flow-matching decoder converts these tokens into waveforms.

We adapt the backbone through two changes (Figure~\ref{fig:conversion}).
First, we replace the causal self-attention mask with a bidirectional one.
Second, we add two block-wise modules, a LoRA and a Conformer-style convolution module, on top of the frozen AR backbone.
We detail these changes in Sections~\ref{sec:bidirectional} and~\ref{sec:adapters}, followed by the training objective in Section~\ref{sec:training} and the inference procedure in Section~\ref{sec:inference}.

\subsection{Bidirectional Adaptation}\label{sec:bidirectional}

Following CosyVoice3's zero-shot input format, each sample is laid out as
\begin{equation}
[\text{SOS}, \mathbf{t}_\text{inst}, \mathbf{t}_\text{prompt}, \mathbf{t}_\text{target}, \text{TASK}, \mathbf{s}_\text{prompt}, \mathbf{s}_\text{target}, \text{EOS}],
\label{eq:input_format}
\end{equation}
where $\mathbf{t}_\text{inst}$, $\mathbf{t}_\text{prompt}$, and $\mathbf{t}_\text{target}$ denote the instruction, prompt, and target text tokens, and $\mathbf{s}_\text{prompt}$ and $\mathbf{s}_\text{target}$ denote the prompt and target speech tokens.
Masking is applied only to $\mathbf{s}_\text{target}$.

During training, a random subset of its positions is replaced with the mask token $\mathrm{M}$, and the model learns to recover the original tokens.
At inference time, $\mathbf{s}_\text{target}$ is initialized as a sequence of $\mathrm{M}$ tokens.

The AR backbone processes the resulting sequence under bidirectional self-attention. 
As a result, each masked target token has direct access to the surrounding speech context, the full transcript, and the prompt speech tokens that carry speaker identity.

\subsection{Lightweight Adapter Architecture}\label{sec:adapters}

We freeze the AR backbone and route all adaptation through two trainable modules attached to every transformer block.
The first module is LoRA~\cite{lora}, which adds low-rank updates to the attention and feed-forward projections of each block while leaving the original weights untouched.
Because the pretrained weights are not modified, the backbone retains its language modeling capabilities and speech-token representations while transitioning from causal to bidirectional attention.
The low-rank updates alone are sufficient to support the newly introduced bidirectional context.
In prior AR-to-dLLM adaptation~\cite{diffullama}, the bidirectional mask must be phased in gradually over training to prevent disruption of the fully trained backbone.
Our approach removes the need for this mask-annealing schedule.

The second module is a Conformer-style convolution module~\cite{conformer}, inserted as a residual branch after each transformer block and composed of a depth-wise convolution, GLU gating, and a Swish activation.
Speech tokens at 25\,Hz exhibit strong temporal locality.
Neighboring positions encode acoustically similar frames, and most of the structure relevant to prosody and phoneme transitions lies within a short window.
Bidirectional self-attention mixes information across the full sequence but has limited inductive bias toward local context.
The convolution module compensates for this limitation with an explicit local-mixing operation, so the two mechanisms jointly cover both global structure and short-range continuity.

\subsection{Training}\label{sec:training}

We follow the masked discrete diffusion formulation~\cite{llada, mdlm}.
Let $\mathbf{s}_0$ denote the original target speech tokens.
For each training example, we sample $t \in (0, 1)$. We then form a partially masked sequence $\mathbf{s}_t$ by replacing each token of $\mathbf{s}_0$ with the mask token $\mathrm{M}$ independently with probability $t$. All other tokens in the input (text, prompt speech, special tokens) remain visible and form the context $\mathbf{c}$.
The model is trained to predict the original token at each masked position:
\begin{equation}\label{eq:loss}
    \mathcal{L}(\theta) \triangleq -\mathbb{E}_{t, \mathbf{s}_0, \mathbf{s}_t}\biggl[\frac{1}{t} \sum_{i=1}^{L} \mathbf{1}[\mathbf{s}_t^i = \mathrm{M}]\, \log p_\theta(\mathbf{s}_0^i \,|\, \mathbf{c}, \mathbf{s}_t)\biggr],
\end{equation}
where $L$ is the target speech length and the indicator restricts the loss to masked positions.
The $1/t$ weighting follows from the ELBO of the masked diffusion process~\cite{diffullama, llada}, amplifying the loss at low masking ratios where the model should predict with high confidence.

The AR backbone predicts the token at position $i{+}1$ from the hidden state at position $i$, whereas standard masked diffusion predicts the masked token at the same position $i$.
We adopt the shift operation from prior AR-to-dLLM conversions~\cite{diffullama, dream}: the hidden state at position $i$ predicts $\mathbf{s}_0^{i+1}$ rather than $\mathbf{s}_0^i$, preserving the AR backbone's next-token prediction behavior.

The mask token $\mathrm{M}$ is the only newly added embedding; we initialize it as the mean of all speech-token embeddings and let training refine it.

\begin{figure*}[t]
\centering
\includegraphics[width=0.9\linewidth]{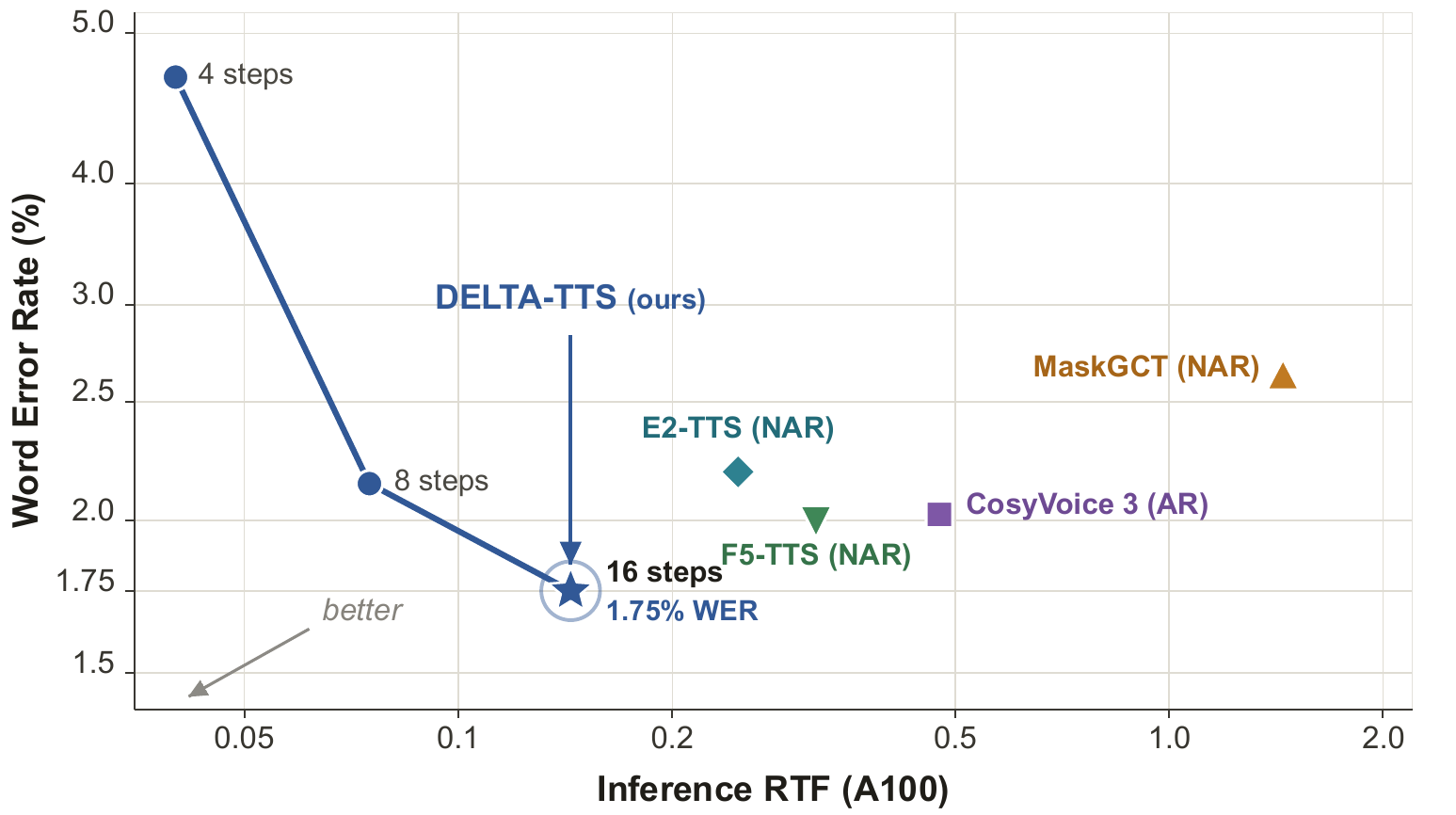}
\caption{Speed--quality Pareto front on Seed-TTS \emph{test-en} (both axes log-scale). DELTA-TTS at 16 steps achieves the lowest WER and RTF among all evaluated systems, offering the most favorable speed--quality trade-off.}
\label{fig:step_tradeoff}
\end{figure*}

\subsection{Inference}\label{sec:inference}

We decode $\mathbf{s}_\text{target}$ through $T$ iterations of parallel unmasking.
At each iteration, the model predicts every masked position simultaneously and draws each candidate with top-$p$ nucleus sampling~\cite{holtzman2019curious}.
We then accept the candidates with the highest confidence, defined as the probability of the sampled token, and keep the rest masked for the next iteration.

Instead of unmasking a uniform number of tokens per step, we use a time-shifted schedule that adapts the continuous formulation of~\cite{sd3} to the discrete unmasking setting.
Let $c_n$ be the cumulative fraction of tokens unmasked after step $n$ out of $T$:
\begin{equation}\label{eq:schedule}
    c_n = \frac{\mu \cdot n/T}{1 + (\mu - 1) \cdot n/T},
\end{equation}
where $\mu > 0$ is a time-shifting parameter, and the number of decisions made at step $n$ is $k_n = \lfloor c_n L \rfloor - \lfloor 
c_{n-1} L \rfloor$.
With $\mu < 1$, the schedule unmasks few tokens at early steps and 
concentrates most decisions toward the end.

This schedule creates a confidence-to-context decoding trajectory. 
When the sequence is still highly masked, the model commits only a few tokens, selecting the most confident positions first. 
Once committed, these tokens become reliable anchors that provide local and bidirectional speech context for the remaining masked positions. 
As decoding proceeds, the accumulated high-confidence tokens help resolve later, less confident positions under richer context.
The time-shifted schedule in Eq.~\eqref{eq:schedule} avoids unreliable early commitments, and the few confident decisions made early support later, more difficult ones.

\section{Experiments}\label{sec:experiments}

\subsection{Experimental Setup}\label{sec:setup}

\subsubsection{Training and Evaluation Datasets}

We train our AR-to-dLLM conversion on LibriTTS~\cite{libritts} (585\,h of English speech).
We evaluate on two zero-shot TTS benchmarks: Seed-TTS \emph{test-en}~\cite{seedtts} (1088 samples) from the Seed-TTS evaluation suite, and the LibriSpeech-PC \emph{test-clean} Subset B~\cite{meister2023librispeech} (1127 samples) released by F5-TTS~\cite{f5tts}.

\subsubsection{Evaluation Metrics}

We report four objective metrics.
For Word Error Rate (WER), which assesses generation intelligibility, we follow prior conventions and use a different ASR system per benchmark: Whisper-Large-V3~\cite{radford2023whisper} for Seed-TTS \emph{test-en} and Faster-Whisper-Large-V3 for LibriSpeech-PC Subset B.
For speaker similarity (SIM), we compute the cosine similarity between the generated and prompt speech using an ECAPA-TDNN model~\cite{desplanques2020ecapa} based on WavLM-large~\cite{wavlm}.
We adopt UTMOS~\cite{utmos}, a predicted mean opinion score, to assess speech naturalness objectively.
We measure inference speed as the Real-Time Factor (RTF) on a single NVIDIA A100.

For subjective evaluation, we collect the comparative mean opinion score (CMOS, $[-3, 3]$) for side-by-side naturalness comparison against human audio, and the similarity mean opinion score (SMOS, $[1, 5]$) for speaker similarity to the prompt audio.

\begin{table*}[t]
\centering
\caption{Objective evaluation on Seed-TTS \emph{test-en}. Baseline results are from official checkpoints or papers. Best results are in \textbf{bold} and second-best are underlined. $\uparrow$: higher is better; $\downarrow$: lower is better.}
\label{tab:main_seedtts}
\begin{tabular*}{\textwidth}{@{\extracolsep{\fill}}clcccc@{}}
\toprule
\multirow{2}{*}{\textbf{Type}} & \multirow{2}{*}{\textbf{Model}} & \multirow{2}{*}{\textbf{Params.}} & \multirow{2}{*}{\textbf{Training Data (hours)}} & \multicolumn{2}{c}{\textbf{Seed-TTS \emph{test-en}}} \\
\cmidrule(lr){5-6}
& & & & \textbf{WER} $\downarrow$ & \textbf{SIM} $\uparrow$ \\
\midrule
-- & \textbf{Ground-truth} & -- & -- & 2.14 & 0.734 \\
\midrule
\multirow{9}{*}{AR}
& Seed-TTS & N/A & N/A & 2.25 & \textbf{0.762} \\
& CosyVoice & 0.3B & 170K Multi. & 4.29 & 0.609 \\
& CosyVoice2 & 0.5B & 170K Multi. & 2.57 & 0.659 \\
& Spark TTS & 0.5B & 100K Multi. & 1.98 & 0.573 \\
& FireRedTTS2 & 1.5B & 1400K Multi. & 1.95 & 0.665 \\
& IndexTTS2 & 1.5B & 55K Emilia & 2.23 & 0.706 \\
& VoxCPM & 0.5B & 1800K Multi. & \underline{1.85} & \underline{0.729} \\
& Llasa & 1B & 250K Multi. & 3.22 & 0.572 \\
& CosyVoice3 & 0.5B & 1000K Multi. & 2.02 & 0.692 \\
\midrule
\multirow{3}{*}{NAR}
& MaskGCT (50 NFE) & 1.1B & 100K Emilia & 2.62 & 0.714 \\
& E2 TTS (32 NFE) & 0.3B & 100K Emilia & 2.19 & 0.710 \\
& F5-TTS (32 NFE) & 0.3B & 100K Emilia & 2.00 & 0.647 \\
\midrule
Ours & DELTA-TTS & 0.5B+94M & 0.585K LibriTTS & \textbf{1.75} & 0.688 \\
\bottomrule
\end{tabular*}
\end{table*}

\begin{table}[t]
\centering
\begin{minipage}[t]{0.40\linewidth}
\centering
\caption{Subjective evaluation on Seed-TTS \emph{test-en}. Means with 95\% confidence intervals.}
\vspace{0.2cm}
\label{tab:subjective}
\small
\setlength{\tabcolsep}{3pt}
\renewcommand{\arraystretch}{1.1}
\resizebox{\linewidth}{!}{
\begin{tabular}{lcc}
\toprule
Model & CMOS $\uparrow$ & SMOS $\uparrow$ \\
\midrule
Ground-truth & 0.00 & $3.80 \pm 0.09$ \\
\midrule
CosyVoice3 & $-0.25 \pm 0.12$ & $4.00 \pm 0.08$ \\
DELTA-TTS & $\mathbf{+0.16} \pm 0.11$ & $\mathbf{4.14} \pm 0.07$ \\
\bottomrule
\end{tabular}}
\end{minipage}
\hspace{0.03\linewidth}
\begin{minipage}[t]{0.54\linewidth}
\centering
\caption{Length-bucketed WER and RTF on Seed-TTS \emph{test-en}. RTF is measured on the token-generation stage.}
\vspace{0.2cm}
\label{tab:length_buckets}
\small
\setlength{\tabcolsep}{3pt}
\resizebox{\linewidth}{!}{
\begin{tabular}{lccccc}
\toprule
\multirow{2}{*}{Duration} & \multicolumn{2}{c}{WER (\%) $\downarrow$} & \multicolumn{2}{c}{RTF $\downarrow$} & \multirow{2}{*}{Speedup $\uparrow$} \\
\cmidrule(lr){2-3} \cmidrule(lr){4-5}
 & CosyVoice3 & DELTA & CosyVoice3 & DELTA & \\
\midrule
0--3\,s     & 1.54 & \textbf{1.24} & 0.479 & \textbf{0.230} & 2.08$\times$ \\
3--5\,s     & 1.74 & \textbf{1.45} & 0.475 & \textbf{0.144} & 3.30$\times$ \\
5--10\,s    & 2.80 & \textbf{2.60} & 0.473 & \textbf{0.106} & 4.46$\times$ \\
\midrule
All         & 2.02 & \textbf{1.75} & 0.475 & \textbf{0.144} & 3.30$\times$ \\
\bottomrule
\end{tabular}}
\end{minipage}
\end{table}

\begin{table*}[t]
\centering
\caption{Objective evaluation on LibriSpeech-PC \emph{test-clean} (Subset B, 1127 samples). Baseline results are from official checkpoints or papers. Best results are in \textbf{bold} and second-best are underlined. $\uparrow$: higher is better; $\downarrow$: lower is better.}
\label{tab:main_librispeech}
\small
\renewcommand{\arraystretch}{1.1}
\resizebox{1.0\linewidth}{!}{
\begin{tabular}{clccccc}
\toprule
\multirow{2}{*}{\textbf{Type}} & \multirow{2}{*}{\textbf{Model}} & \multirow{2}{*}{\textbf{Params.}} & \multirow{2}{*}{\textbf{Training Data (hours)}} & \multicolumn{3}{c}{\textbf{LibriSpeech-PC \emph{test-clean}}} \\
\cmidrule(lr){5-7}
& & & & \textbf{WER} $\downarrow$ & \textbf{SIM} $\uparrow$ & \textbf{UTMOS} $\uparrow$ \\
\midrule
-- & \textbf{Ground-truth} & -- & -- & 2.23 & 0.69 & 4.10 \\
\midrule
\multirow{4}{*}{AR}
& CosyVoice & 0.3B & 170K Multi. & 3.59 & 0.66 & 4.14 \\
& FireRedTTS & 0.6B & 248K Multi. & 2.69 & 0.47 & -- \\
& DiTAR & 0.6B & 100K Emilia & 2.39 & \underline{0.67} & 4.22 \\
& CosyVoice3 & 0.5B & 1000K Multi. & \textbf{1.95} & \textbf{0.69} & \textbf{4.28} \\
\midrule
\multirow{3}{*}{NAR}
& MaskGCT (50 NFE) & 1.1B & 100K Emilia & 2.72 & \textbf{0.69} & 3.90 \\
& E2 TTS (32 NFE) & 0.3B & 100K Emilia & 2.95 & \textbf{0.69} & 3.56 \\
& F5-TTS (32 NFE) & 0.3B & 100K Emilia & 2.42 & 0.66 & 3.88 \\
\midrule
Ours & DELTA-TTS & 0.5B+94M & 0.585K LibriTTS & \underline{2.17} & \textbf{0.69} & \underline{4.27} \\
\bottomrule
\end{tabular}}
\end{table*}

\subsubsection{Baselines}

We compare \textbf{DELTA-TTS} against representative AR and NAR zero-shot TTS systems.
AR baselines include Seed-TTS~\cite{seedtts}, CosyVoice~\cite{du2024cosyvoice}, CosyVoice2~\cite{cosyvoice2}, Spark TTS~\cite{sparktts}, FireRedTTS~\cite{guo2025fireredtts}, FireRedTTS2~\cite{fireredtts2}, IndexTTS2~\cite{indextts2}, Llasa~\cite{llasa}, VoxCPM~\cite{zhou2025voxcpm}, DiTAR~\cite{ditar}, and our base model CosyVoice3~\cite{cosyvoice3}; NAR baselines include MaskGCT~\cite{maskgct}, E2 TTS~\cite{e2tts}, and F5-TTS~\cite{f5tts}.
For Seed-TTS \emph{test-en}, baseline numbers are taken from the CosyVoice HuggingFace evaluation page\footnote{\url{https://huggingface.co/FunAudioLLM/CosyVoice-300M\#evaluation}} where available, and from the Llasa paper~\cite{llasa} for E2 TTS and Llasa.

For LibriSpeech-PC Subset B, baseline numbers are taken from F5-TTS~\cite{f5tts} and DiTAR~\cite{ditar} where available, or measured by us using the released official checkpoints.
For CosyVoice3 in particular, we measure WER, SIM, and UTMOS ourselves on both benchmarks using the released checkpoint, to ensure consistency with the rest of the evaluation pipeline.

\subsection{Main Results}\label{sec:main_results}

\subsubsection{Intelligibility}

On Seed-TTS \emph{test-en} (Table~\ref{tab:main_seedtts}), DELTA-TTS achieves the lowest WER among all compared systems despite training on only 585\,h of English speech.
On LibriSpeech-PC Subset B (Table~\ref{tab:main_librispeech}), DELTA-TTS again achieves a lower WER than every NAR baseline and is second only to the strongest AR system, showing that the conversion preserves the synthesis capability of the pretrained backbone.
These results are obtained by training only $94$M parameters on top of the frozen backbone, without retraining the AR weights.

\subsubsection{Speaker Similarity and Naturalness}

DELTA-TTS achieves SIM competitive with leading AR and NAR baselines on both benchmarks, and UTMOS on Subset B is on par with the strongest AR systems, indicating that the parallel paradigm does not degrade perceptual audio quality.
Subjective evaluation in Table~\ref{tab:subjective} reinforces this finding. DELTA-TTS is judged more natural than ground truth in CMOS ($+0.16$, against $-0.25$ for CosyVoice3) and obtains the highest SMOS ($4.14$ vs.\ $4.00$ for CosyVoice3 and $3.80$ for ground truth), with both gaps over CosyVoice3 statistically significant (Appendix~\ref{sec:appendix_subjective}).

\subsection{Speed--Quality Trade-off}\label{sec:speed_quality}
Figure~\ref{fig:step_tradeoff} reports the speed--quality Pareto front on Seed-TTS \emph{test-en}, obtained by sweeping the number of unmasking steps over $\{4, 8, 16\}$.
With too few steps the masked positions are under-resolved and WER rises sharply, while quality improves steadily as the step budget grows.
At $16$ steps, DELTA-TTS reaches the bottom-left corner of the front, and we adopt this setting as our default.
DELTA-TTS then achieves a WER of $1.75\%$ at an RTF of $0.144$, a $3.3\times$ speedup over the AR baseline, and dominates the NAR baselines on both axes.
RTF here is measured on the token-generation stage alone; end-to-end latency including the flow-matching vocoder is reported in Appendix~\ref{sec:appendix_latency}, where the speedup is $2.71\times$ at batch $1$.

Table~\ref{tab:length_buckets} further shows that DELTA-TTS's fixed unmasking budget produces a larger speedup on longer utterances ($2.08\times \to 4.46\times$ from $<3$\,s to $\geq 5$\,s) while retaining the WER advantage at every length.

\subsection{Ablation Study}\label{sec:ablation}

Table~\ref{tab:train_ablation} breaks down each component of our adaptation. Adding the time-shifted schedule reduces WER over the uniform-schedule baseline, confirming the benefit of deferring uncertain commitments to later steps.
The block-wise convolution module further reduces WER by introducing short-range mixing that complements bidirectional attention. Raising the LoRA rank to a comparable trainable budget without the convolution reduces WER only to $2.13\%$, so the gain comes from the convolutional inductive bias rather than added capacity.
Prompt-conditioned training trades a small amount of WER and naturalness for a substantial gain in speaker similarity.

Fully unfreezing the backbone increases WER despite more parameters, suggesting that the limited 585-hour adaptation data is insufficient to fine-tune all 494M parameters without overfitting. Parameter-efficient adapters are therefore preferable in this low-resource setting.
Fine-tuning the AR backbone on the same corpus with LoRA while keeping causal decoding reaches $1.87\%$, so part of the gain comes from English-only adaptation; the conversion accounts for the remainder as well as the entire speedup.
A knowledge-distillation baseline that trains the same dLLM to match CosyVoice3's outputs lags our direct conversion in both WER and SIM, showing that distilling from a teacher caps the student at the teacher's quality, whereas direct conversion can surpass the AR backbone.

For the rule-based length variant, we use the length estimation rule of OmniVoice~\cite{zhu2026omnivoice}, which sets the target speech-token count from the audio-to-character rate of the prompt and the weighted character count of the target text.

\begin{table*}[t]
\centering
\caption{Ablation of DELTA-TTS on Seed-TTS \emph{test-en}. Rows marked ``+'' are cumulative on top of the preceding row. The two controls isolate trainable capacity and English-only adaptation, respectively, from the conversion itself.}
\label{tab:train_ablation}
\small
\renewcommand{\arraystretch}{1.15}
\resizebox{1.0\linewidth}{!}{
\begin{tabular}{lcccc}
\toprule
\textbf{Configuration} & \textbf{WER} (\%) $\downarrow$ & \textbf{SIM} $\uparrow$ & \textbf{UTMOS} $\uparrow$ & \textbf{Trainable} \\
\midrule
\multicolumn{5}{l}{\emph{Cumulative adaptation}} \\
Naive conversion (LoRA, no Conv, uniform sched.) & 3.01 & 0.574 & 4.04 & 35M \\
\quad + Time-shifted schedule                    & 2.59 & 0.586 & 4.05 & 35M \\
\quad + Convolution module ($k{=}31$)            & \textbf{1.61} & 0.584 & \textbf{4.09} & 94M \\
\quad + Prompt conditioning (GT length)          & 1.63 & 0.686 & 4.01 & 94M \\
\quad + Prompt conditioning (rule-based length)  & 1.75 & \textbf{0.688} & 3.97 & 94M \\
\midrule
\multicolumn{5}{l}{\emph{Capacity-matched control}} \\
LoRA only ($r{=}160$, no Conv)                   & 2.13 & 0.576 & 4.02 & 88M \\
\midrule
\multicolumn{5}{l}{\emph{Control: adaptation without conversion}} \\
LoRA AR fine-tuning (causal)                     & 1.87 & 0.687 & 3.98 & 35M \\
\midrule
\multicolumn{5}{l}{\emph{Alternative training strategies}} \\
Full fine-tuning                                 & 1.97 & 0.680 & 4.01 & 494M \\
KD from CosyVoice3                               & 2.37 & 0.591 & 4.05 & 94M \\
\bottomrule
\end{tabular}}
\end{table*}

\subsection{Backbone Transfer}\label{sec:backbone_transfer}

The ablations above are all measured on a single backbone, so we test whether the conversion recipe itself is backbone-agnostic by applying it to two additional AR TTS systems, Llasa (1B) and CosyVoice2, keeping the same LoRA rank, the same per-block convolution, and the same LibriTTS adaptation data.
For Llasa, whose codec produces $50$\,Hz tokens and therefore roughly twice the sequence length of CosyVoice3, we adjust only the decoding schedule to $32$ unmasking steps; CosyVoice2 uses our CosyVoice3 configuration unchanged.

\begin{table}[t]
\centering
\caption{Backbone transfer on Seed-TTS \emph{test-en}. The same adapter recipe is applied to two additional AR backbones without further tuning.}
\vspace{0.2cm}
\label{tab:backbone_transfer}
\small
\renewcommand{\arraystretch}{1.15}
\begin{tabular}{lcc}
\toprule
\textbf{Model} & \textbf{WER} (\%) $\downarrow$ & \textbf{SIM} $\uparrow$ \\
\midrule
Llasa (AR backbone) & 3.72 & \textbf{0.519} \\
\quad + DELTA-TTS & \textbf{3.13} & 0.454 \\
\midrule
CosyVoice2 (AR backbone) & 2.50 & 0.602 \\
\quad + DELTA-TTS & \textbf{2.31} & \textbf{0.614} \\
\bottomrule
\end{tabular}
\end{table}

The conversion improves WER on both backbones (Table~\ref{tab:backbone_transfer}), and on CosyVoice2 it improves speaker similarity as well, indicating that the recipe does not depend on the particular tokenizer or backbone it was developed on.

\subsection{Analysis}\label{sec:analysis}

\begin{figure*}[t]
\centering
\subfloat[Text--speech alignment]{%
    \includegraphics[width=0.55\linewidth]{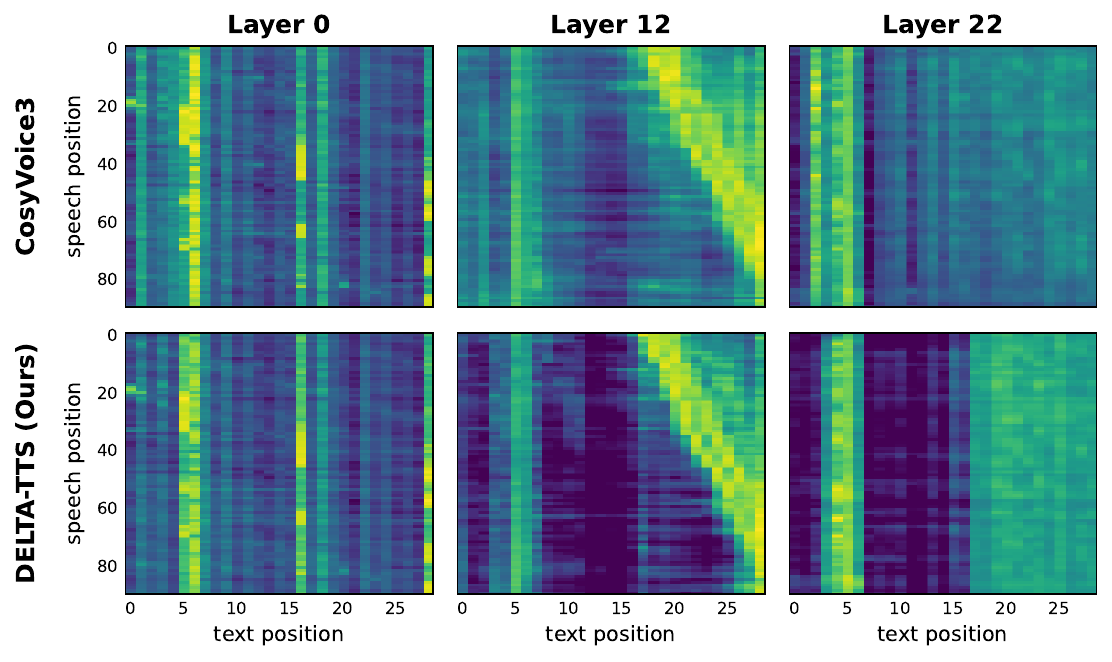}%
    \label{fig:attention_alignment}}
\hfill
\subfloat[Attention budget of masked tokens]{%
    \includegraphics[width=0.45\linewidth]{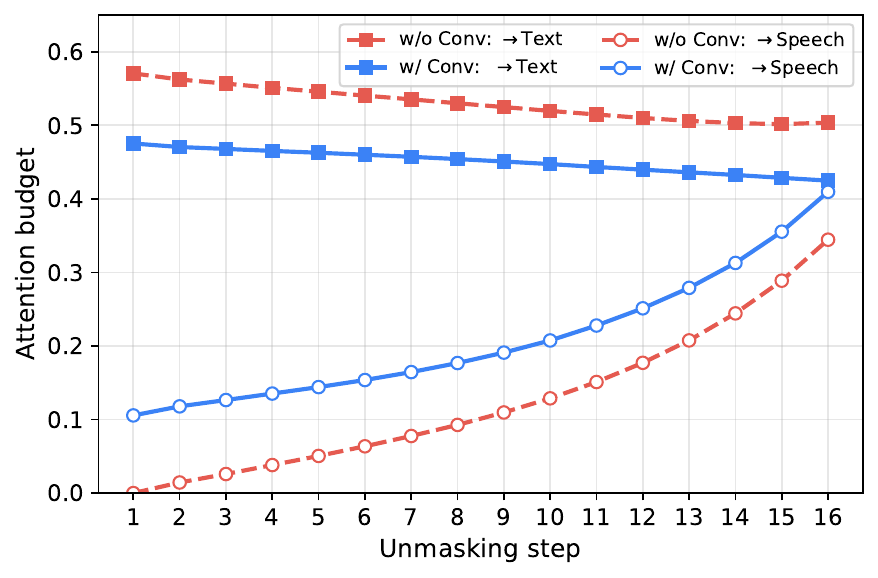}%
    \label{fig:conv_effects}}
\caption{(a) Text--speech alignment across depth: DELTA-TTS sharpens the middle-layer diagonal and, in the late layer, attends to upcoming text, a pattern unavailable to the causal AR baseline. (b) Per-step attention budget from masked tokens; the convolution module shifts more attention onto neighboring unmasked speech while keeping text attention stable.}
\label{fig:analysis}
\end{figure*}

\begin{figure*}[t]
\centering
\subfloat[AR confidence by token position]{%
    \includegraphics[width=0.5\linewidth]{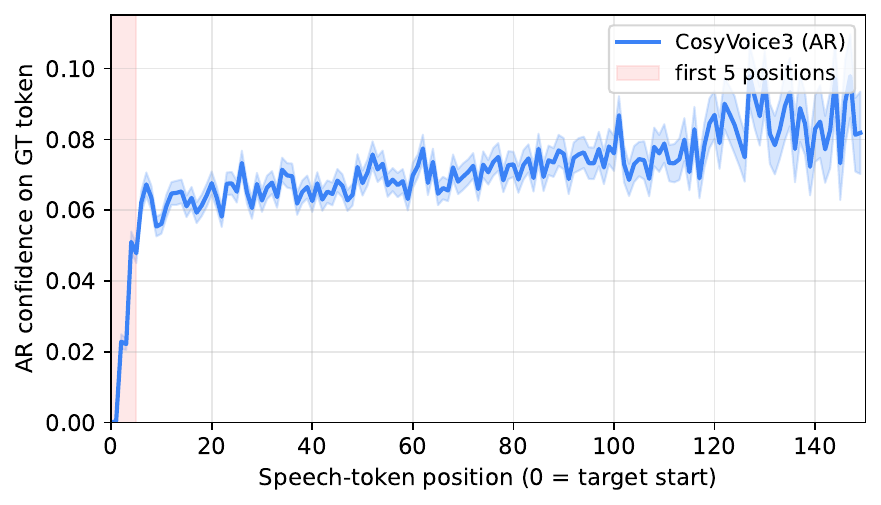}%
    \label{fig:conf_position}}
\hfill
\subfloat[Marginal confidence distribution]{%
    \includegraphics[width=0.5\linewidth]{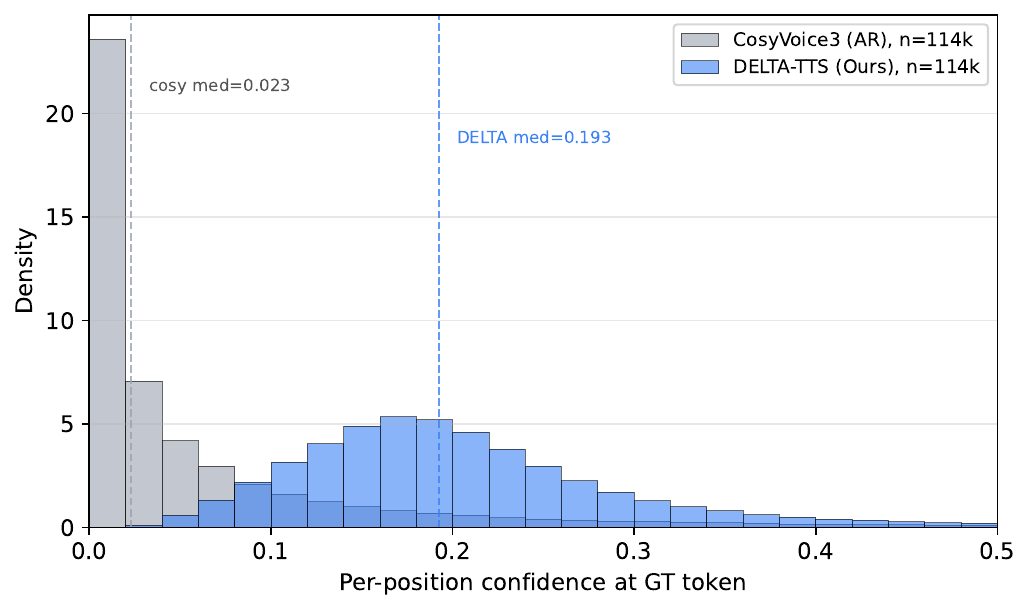}%
    \label{fig:conf_per_step}}
\caption{(a) Mean teacher-forced AR confidence at the GT speech token across $1088$ Seed-TTS \emph{test-en} utterances; the first $5$ positions sit close to zero. (b) Marginal distribution of per-position confidence at the GT speech token; the AR model's teacher-forced confidence concentrates near zero, while DELTA-TTS commits at far higher confidence.}
\label{fig:confidence_analysis}
\end{figure*}

We analyze DELTA-TTS from two angles.
We first examine how the conversion reshapes attention over text and speech, and then how confidence-ordered decoding mitigates the hallucinations that afflict AR generation.

\subsubsection{Text--Speech Alignment and Speech-Token Locality}
Figure~\ref{fig:analysis}(a) shows that DELTA-TTS produces a substantially cleaner text--speech alignment than CosyVoice3, with the middle-layer diagonal sharpening from a noisy blur into a clearly defined line.
This sharpening does not come from the adaptation data but from the convolution module: a LoRA fine-tune on the same corpus with causal attention does not sharpen at all (Appendix~\ref{sec:appendix_sharpness}).

Figure~\ref{fig:analysis}(b) shows that masked positions concentrate more attention on neighboring unmasked speech tokens while attention to text remains stable, and this strengthened local focus tightens the alignment.
Beyond this, the late layer attends to upcoming text, exploiting future context that is structurally inaccessible to the causal AR decoder.

\subsubsection{Confidence Advantage and Robustness to AR Failure Modes}\label{sec:why_dllm}
Autoregressive decoders must commit each token before observing the tokens that follow, so early positions are decided under insufficient context and the resulting errors propagate.
DELTA-TTS removes this constraint.
Bidirectional attention grants every position access to future context, and confidence-ordered decoding defers the positions that require more context until enough has accumulated.

Figure~\ref{fig:confidence_analysis}(a) plots the AR model's confidence at the ground-truth (GT) speech token at each position.
The first 5 positions sit close to zero: the AR model places almost no probability mass on the GT token at the utterance start, yet must commit these positions from the prompt context alone.
DELTA-TTS instead defers them through confidence-ordered decoding (Appendix~\ref{appendix:conf_dist}).
Figure~\ref{fig:confidence_analysis}(b) aggregates these per-position confidences into a marginal distribution. 
The AR model's teacher-forced confidence concentrates near zero (median $0.023$), whereas DELTA-TTS commits at far higher confidence (median $0.193$, $8\times$ higher), with negligible mass in that region.

This confidence advantage has a measurable consequence.
Over the $5$ random seeds of our sampling stability sweep (Appendix~\ref{app:sampling}), we examine the tail of the per-utterance WER distribution across all $5440$ generations per system (Table~\ref{tab:tail_wer}).
The AR backbone collapses completely on 5 generations, abandoning the target text, while DELTA-TTS never does and its worst case remains a partial error at $64\%$.
The error decomposition points the same way: insertions, the signature of repetition-style hallucination, drop by $47.9\%$ (Appendix~\ref{sec:appendix_errortypes}).

\begin{table}[t]
\centering
\caption{Tail of the per-utterance WER distribution on Seed-TTS \emph{test-en} over $5440$ generations per system ($5$ seeds $\times$ $1088$ utterances). The AR backbone abandons the target text entirely on 5 generations, whereas DELTA-TTS never does.}
\vspace{0.2cm}
\label{tab:tail_wer}
\small
\renewcommand{\arraystretch}{1.15}
\begin{tabular}{lcc}
\toprule
& CosyVoice3 (AR) & DELTA-TTS \\
\midrule
Complete collapse (WER $\geq 80\%$) & $5$ & $\mathbf{0}$ \\
WER $\geq 50\%$                     & $7$ & $\mathbf{3}$ \\
Worst-case per-utterance WER        & $100\%$ & $\mathbf{64\%}$ \\
Perfect transcription (WER $=0$)    & $83.9 \pm 0.3\%$ & $\mathbf{86.2 \pm 0.8\%}$ \\
\bottomrule
\end{tabular}
\end{table}

\section{Conclusion}\label{sec:conclusion}

We presented DELTA-TTS, a lightweight framework that converts a pretrained autoregressive TTS model into a discrete diffusion language model with only about $15\%$ added parameters.
On top of LoRA, a block-wise convolution module injected the local acoustic context that bidirectional attention captures weakly, and a $1/t$-weighted objective with a time-shifted schedule deferred low-confidence positions to later steps.
Trained on only $585$ hours of LibriTTS, DELTA-TTS achieved a $1.75\%$ WER on Seed-TTS \emph{test-en}, improving on the $2.02\%$ of its AR backbone while generating tokens $3.3\times$ faster.
We attributed these gains to local mixing and to confidence-ordered decoding, which deferred the positions where the AR model is least confident until sufficient context has accumulated and therefore reduced hallucination and error propagation.
These results support our hypothesis that confidence-ordered decoding produces fewer catastrophic failures than strictly left-to-right decoding.

\begin{ack}  
This work was partly supported by the National Research Foundation of Korea (NRF) grant funded by the Korea government (MSIT) (RS-2026-25602529). This work was also partly supported by the Institute of Information \& Communications Technology Planning \& Evaluation (IITP) grants funded by the Korea government (MSIT) (IITP-2026-RS-2026-25615334, AI Star Fellowship Support Program; RS-2019-II190421, Artificial Intelligence Graduate School Program (Sungkyunkwan University)). This work was also partly supported by the Institute of Information \& Communications Technology Planning \& Evaluation (IITP, AI Computing Support Project for R\&D) grant funded by the Korea government (MSIT) (High-Performance Research AI Computing Infrastructure Support at the 2 PFLOPS Scale, 25\%).
\end{ack}
\newpage
{
    \small
    \bibliographystyle{plain}
    \bibliography{references}
}


\newpage
\appendix
\section{Appendix}

\subsection{Implementation Details}\label{sec:appendix_impl}

We train on a single NVIDIA A100 with bfloat16 mixed precision. The LoRA targets the query, key, value, output, gate, up, and down projections of every transformer block, with rank $r{=}64$ and scaling $\alpha{=}128$. The depth-wise convolution uses kernel size $31$ and dropout $0.1$. We optimize with AdamW at a constant learning rate of $1{\times}10^{-4}$ after a $2000$-step linear warmup, with an effective batch size of $16$ utterances obtained via gradient accumulation.

\subsection{End-to-End Latency}\label{sec:appendix_latency}

\begin{table}[h]
\centering
\caption{End-to-end RTF including the flow-matching vocoder, on Seed-TTS \emph{test-en}.}
\vspace{0.2cm}
\label{tab:e2e_latency}
\small
\renewcommand{\arraystretch}{1.15}
\begin{tabular}{lccccc}
\toprule
\multirow{2}{*}{Batch} & CosyVoice3 (AR) & \multicolumn{2}{c}{DELTA-TTS ($8$ steps)} & \multicolumn{2}{c}{DELTA-TTS ($16$ steps)} \\
\cmidrule(lr){2-2} \cmidrule(lr){3-4} \cmidrule(lr){5-6}
 & RTF $\downarrow$ & RTF $\downarrow$ & Speedup $\uparrow$ & RTF $\downarrow$ & Speedup $\uparrow$ \\
\midrule
$1$ & $0.5967$ & $0.1453$ & $4.11\times$ & $0.2201$ & $2.71\times$ \\
$2$ & $0.3403$ & $0.0760$ & $4.48\times$ & $0.1081$ & $3.15\times$ \\
$4$ & $0.2108$ & $0.0421$ & $\mathbf{5.01\times}$ & $0.0612$ & $\mathbf{3.44\times}$ \\
$8$ & $0.1343$ & $0.0297$ & $4.52\times$ & $0.0401$ & $3.35\times$ \\
\bottomrule
\end{tabular}
\end{table}

We measure RTF over the full generation pipeline, including text processing, speech-token generation, and the flow-matching vocoder, so that the reported numbers reflect user-visible latency rather than the token-generation stage alone.
All measurements use a single NVIDIA A100. CosyVoice3 runs its official decoder with KV caching enabled, and DELTA-TTS is measured at $8$ and $16$ unmasking steps.
RTF is wall-clock time divided by generated-audio duration, averaged over Seed-TTS \emph{test-en}.

The speedup holds across batch sizes and peaks at batch $4$. It is smaller than the speedup measured on the token-generation stage alone at batch 1, since the shared flow-matching vocoder is unaffected by the conversion and takes a fixed share of the wall-clock time in both systems.

\subsection{Alignment Sharpness Control}\label{sec:appendix_sharpness}

\begin{table}[h]
\centering
\caption{Alignment sharpness control on Seed-TTS \emph{test-en}. Sharpness is the fraction of text-to-speech attention mass falling within $\pm 1$ text token of the position identified by MFA forced alignment, taken from each utterance's best-aligned head.}
\vspace{0.2cm}
\label{tab:sharpness_control}
\small
\setlength{\tabcolsep}{6pt}
\renewcommand{\arraystretch}{1.15}
\begin{tabular}{lccccc}
\toprule
Model & WER (\%) $\downarrow$ & SIM $\uparrow$ & UTMOS $\uparrow$ & Sharpness $\uparrow$ & Relative \\
\midrule
CosyVoice3 (backbone)            & 2.02 & \textbf{0.692} & 3.955 & 0.725 & -- \\
\quad + LoRA fine-tune (control) & 1.87 & 0.687 & \textbf{3.982} & 0.720 & $-0.7\%$ \\
DELTA-TTS                        & \textbf{1.75} & 0.688 & 3.975 & \textbf{0.788} & $+8.8\%$ \\
\bottomrule
\end{tabular}
\end{table}

The sharper alignment in Figure~\ref{fig:analysis}(a) could in principle come from fine-tuning on a narrow domain rather than from the conversion itself.
We therefore train a control that keeps attention causal, using the identical LoRA configuration, the same LibriTTS data, and the official CosyVoice3 objective.
We then measure diagonal alignment sharpness on Seed-TTS \emph{test-en}.

Fine-tuning on LibriTTS improves the backbone's WER from $2.02\%$ to $1.87\%$. The control's attention nonetheless does not sharpen, while DELTA-TTS is $8.8\%$ sharper, indicating that the sharper alignment does not follow from training on this data.

\subsection{Expressivity Preservation}\label{sec:appendix_expressivity}

\begin{table}[h]
\centering
\caption{Emotion cloning on the CV3-eval emotion subset (en, $150$ utterances) with happy, sad, and angry prompts. Spread is the gap between the extreme emotions; parentheses give DELTA-TTS's spread as a fraction of the backbone's.}
\vspace{0.2cm}

\label{tab:cv3eval_emotion}
\small
\setlength{\tabcolsep}{6pt}
\renewcommand{\arraystretch}{1.15}
\begin{tabular}{llcccc}
\toprule
Metric & Model & sad & happy & angry & Spread \\
\midrule
\multirow{2}{*}{Median F0 (Hz)}
 & CosyVoice3 & 162.2 & 182.3 & 205.7 & 43.5 \\
 & DELTA-TTS  & 162.6 & 172.7 & 200.6 & 38.1 ($88\%$) \\
\midrule
\multirow{2}{*}{F0 variability (st)}
 & CosyVoice3 & 4.38 & 4.57 & 5.04 & 0.67 \\
 & DELTA-TTS  & 4.54 & 4.71 & 5.24 & 0.69 ($104\%$) \\
\midrule
\multirow{2}{*}{Speaking rate (char/s)}
 & CosyVoice3 & 12.9 & 13.9 & 15.4 & 2.5 \\
 & DELTA-TTS  & 16.0 & 17.9 & 18.3 & 2.3 ($92\%$) \\
\bottomrule
\end{tabular}
\end{table}

\begin{table}[h]
\centering
\caption{Expressive continuation on the CV3-eval subjective-continuation subset (en, $90$ utterances). Values are the correlation between the prompt's attribute and the generation's, with bootstrap $95\%$ intervals.}
\vspace{0.2cm}
\label{tab:cv3eval_continuation}
\small
\setlength{\tabcolsep}{6pt}
\renewcommand{\arraystretch}{1.15}
\begin{tabular}{lcc}
\toprule
Axis (prompt $\rightarrow$ generation) & CosyVoice3 & DELTA-TTS \\
\midrule
Speaking rate                    & $+0.94$ $[0.88, 0.98]$ & $+0.87$ $[0.76, 0.95]$ \\
Volume (RMS dB)                  & $+0.98$ $[0.96, 0.99]$ & $+0.93$ $[0.87, 0.97]$ \\
Emotion (F0 variability, st)     & $+0.76$ $[0.63, 0.88]$ & $+0.55$ $[0.29, 0.73]$ \\
Rhythm (F0 variability, st)      & $+0.76$ $[0.37, 0.92]$ & $+0.70$ $[0.25, 0.88]$ \\
\midrule
Pooled ($90$, z-scored)          & $+0.85$ $[0.78, 0.91]$ & $+0.74$ $[0.64, 0.82]$ \\
\bottomrule
\end{tabular}
\end{table}

The adapter is trained on LibriTTS, a corpus of read audiobook speech with little emotional or rhythmic variation, so we test whether the conversion preserves the expressive capability of the multilingual backbone. We evaluate on the two expressive subsets of CV3-eval, measuring CosyVoice3 and DELTA-TTS under the same pipeline.

DELTA-TTS preserves the emotion ordering on every metric (sad $<$ happy $<$ angry) and retains $88$--$104\%$ of the backbone's separation between emotions. The speaking rate is uniformly faster, with total, voiced, and silence durations all about $80\%$ of the backbone's, which we attribute to the rule-based length estimator (Section~\ref{sec:ablation}): the short prompts in CV3-eval make the target length harder to estimate.

The prompt's expressive attributes are carried into the generation rather than flattened, most strongly for speed and volume. The adapter was never exposed to expressive prompts of this kind during adaptation, so the emotional and prosodic conditioning measured in both tables appears to be inherited from the backbone rather than learned from the training data.

\subsection{Sampling Stability Sweep}\label{app:sampling}

\begin{table}[h]
\centering
\caption{Sampling stability on Seed-TTS \emph{test-en}. Stochastic results are mean$\pm$std over $5$ random seeds.}
\vspace{0.2cm}
\label{tab:sampling_stability}
\begin{tabular}{lccc}
\toprule
Configuration & WER (\%) $\downarrow$ & SIM $\uparrow$ & UTMOS $\uparrow$ \\
\midrule
DELTA-TTS, stochastic                   & $\mathbf{1.75 \pm 0.10}$ & $0.688 \pm 0.001$    & $\mathbf{3.975 \pm 0.005}$ \\
\midrule
CosyVoice3, stochastic                  & $1.99 \pm 0.09$          & $\mathbf{0.692 \pm 0.001}$    & $3.955 \pm 0.004$ \\
\bottomrule
\end{tabular}
\end{table}

We evaluate sampling stability by running each model with $5$ different random seeds on Seed-TTS \emph{test-en} ($1088$ utterances) under the default configuration (DELTA-TTS: $\mu = 0.3$). Since Seed-TTS provides no dev split, $\mu$ was selected on a held-out LibriTTS dev-clean set rather than on the evaluation set. We set $\text{top\_p}{=}0.8$ for both models, following CosyVoice3's official sampling configuration.\footnote{\url{https://github.com/FunAudioLLM/CosyVoice}}

\subsection{Error-Type Decomposition}\label{sec:appendix_errortypes}

\begin{table}[h]
\centering
\caption{WER decomposed into substitutions, insertions, and deletions on Seed-TTS \emph{test-en}, as a percentage of reference words.}
\vspace{0.2cm}
\label{tab:error_types}
\small
\setlength{\tabcolsep}{8pt}
\renewcommand{\arraystretch}{1.15}
\begin{tabular}{lccc}
\toprule
Error type (\% of reference words) & CosyVoice3 (AR) & DELTA-TTS & Relative change \\
\midrule
Substitution & $1.559$ & $\mathbf{1.313}$ & $-15.8\%$ \\
Insertion    & $0.117$ & $\mathbf{0.061}$ & $-47.9\%$ \\
Deletion     & $\mathbf{0.315}$ & $0.351$ & $+11.4\%$ \\
\bottomrule
\end{tabular}
\end{table}

We decompose the WER on Seed-TTS \emph{test-en} into substitutions, insertions, and deletions. Insertions, the signature of repetition-style hallucination, drop by $47.9\%$ and substitutions by $15.8\%$, while deletions are marginally higher (Table~\ref{tab:error_types}).

\subsection{Subjective Evaluation Details}\label{sec:appendix_subjective}

\begin{table}[h]
\centering
\caption{A/B preference test on $100$ utterances drawn independently of the MOS set.}
\vspace{0.2cm}
\label{tab:preference}
\small
\renewcommand{\arraystretch}{1.15}
\begin{tabular}{lccc}
\toprule
& CosyVoice3 & Neutral & DELTA-TTS \\
\midrule
Overall quality    & $21\%$ & $43\%$ & $\mathbf{36\%}$ \\
Speaker similarity & $22\%$ & $49\%$ & $\mathbf{29\%}$ \\
\bottomrule
\end{tabular}
\end{table}

For the subjective evaluation in Table~\ref{tab:subjective}, we recruited $30$ adult volunteers anonymously from the authors' institution, each rating $100$ samples per system from Seed-TTS \emph{test-en}. Listeners rated SMOS (speaker similarity to the prompt voice, $1$--$5$) for all three systems and CMOS (naturalness against ground truth, $-3$ to $+3$) for the two synthetic ones. DELTA-TTS obtains significantly higher CMOS and SMOS than CosyVoice3 (Wilcoxon signed-rank test; $p < 0.001$ and $p < 0.01$).

We additionally run an A/B preference test with the same raters on $100$ utterances drawn independently of the MOS set. Excluding neutral choices, DELTA-TTS is preferred significantly more often on both axes (Table~\ref{tab:preference}; binomial test; $p < 0.001$ for quality, $p < 0.01$ for speaker similarity).

In our SMOS setup, raters hear the reference prompt and rate all systems against it. Since the ground truth is a different utterance by the same speaker, it naturally differs from the reference in style and recording conditions, while the TTS systems explicitly target that reference. Synthetic samples scoring above the ground truth are therefore an expected outcome of this protocol.

\subsection{Time-Shifted Schedule and \texorpdfstring{$1/t$}{1/t} Loss}\label{app:tshift}

\begin{figure}[h]
\centering
\includegraphics[width=1.0\linewidth]{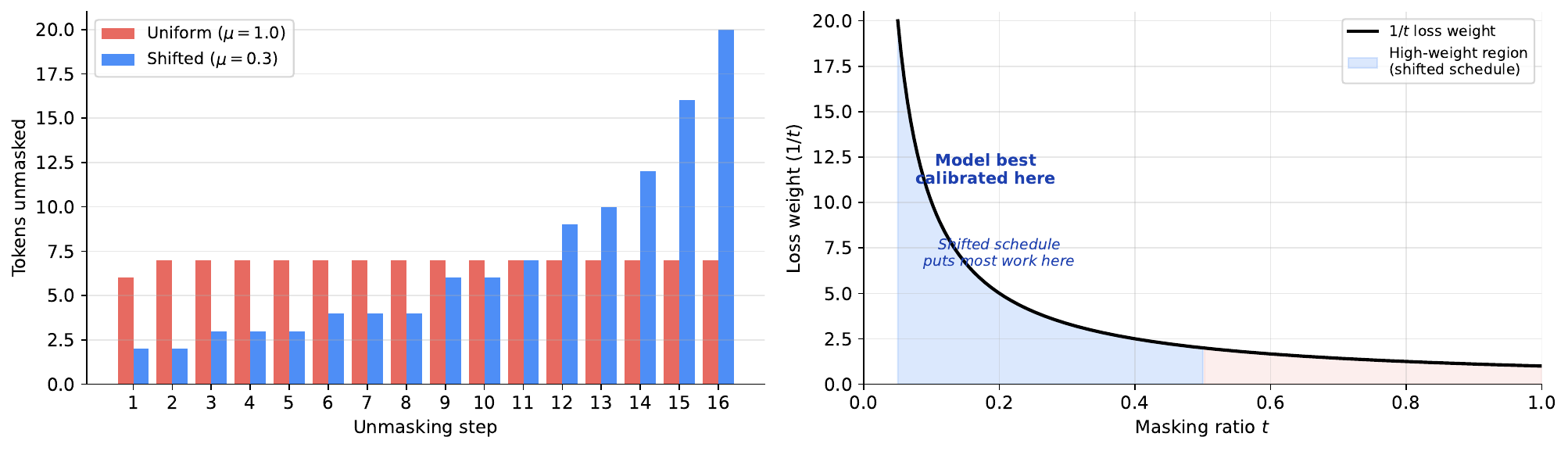}
\caption{Number of tokens unmasked per step under uniform and shifted schedules (left), and the $1/t$ loss weight as a function of the masking ratio (right).}
\label{fig:tshift_synergy}
\end{figure}

We use a time-shifted schedule that defers most unmasking decisions to the later steps of decoding.
As shown in Figure~\ref{fig:tshift_synergy}, the shifted schedule concentrates these decisions at low masking ratios, where the $1/t$ loss places its largest weights.

\subsection{Easy-first, Hard-last Unmasking Order}\label{app:easy_hard}

\begin{figure}[h]
\centering
\includegraphics[width=0.8\linewidth]{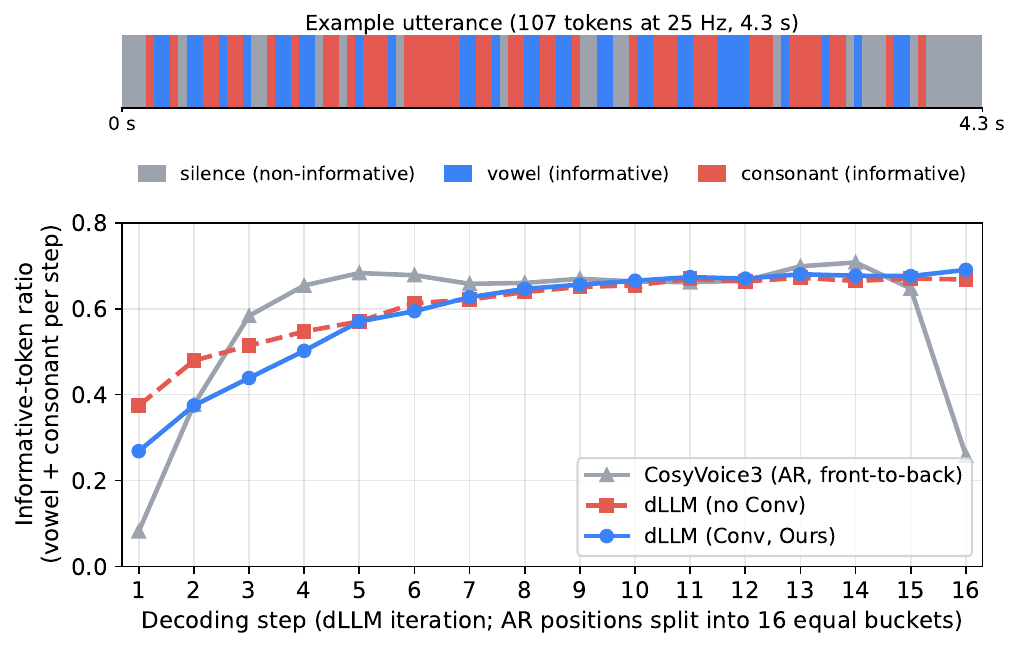}
\caption{Per-step informative-token (vowel\,+\,consonant) ratio averaged over 1088 utterances. Strip on top: phoneme classes on one example utterance.}
\label{fig:easy_hard_informative}
\end{figure}

Figure~\ref{fig:easy_hard_informative} plots the per-step informative-token (vowel\,+\,consonant) ratio.
Under the AR order (CosyVoice3 audio aligned via MFA), the ratio peaks at the middle steps and drops at both ends, since the AR decoder commits the leading and trailing silence at the first and last steps.
Under DELTA-TTS, the ratio increases monotonically: the model commits silence first and defers phonetically meaningful tokens until enough bidirectional context has accumulated.
This phoneme-class deferral mirrors the AR-difficulty deferral in Figure~\ref{fig:commit_ordering}.

\subsection{Per-Step Commit Ordering}\label{appendix:conf_dist}

\begin{figure}[h]
\centering
\includegraphics[width=0.7\linewidth]{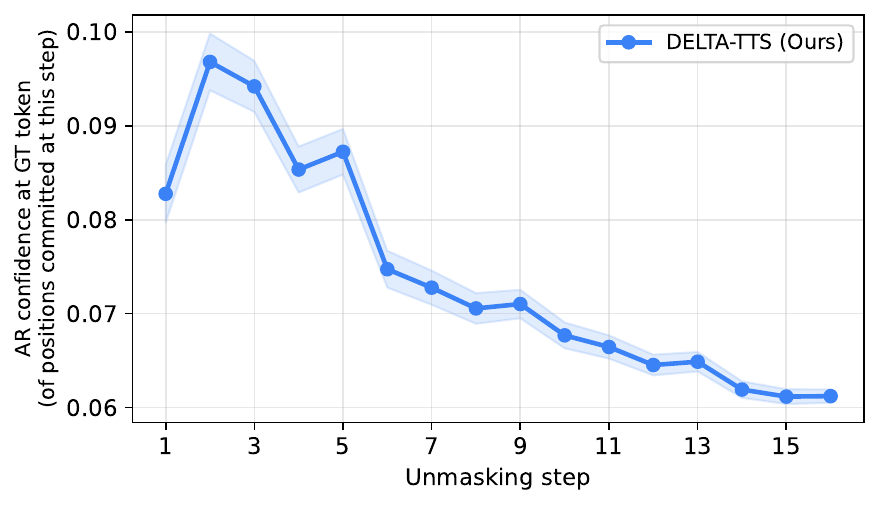}
\caption{Mean teacher-forced AR confidence (GT token) of the positions DELTA-TTS commits at each unmasking step, over $1088$ utterances. AR-low-confidence positions are deferred to later steps.}
\label{fig:commit_ordering}
\end{figure}

Figure~\ref{fig:commit_ordering} shows, at each unmasking step, the mean teacher-forced AR confidence of the positions DELTA-TTS commits. DELTA-TTS commits high-AR-confidence positions early and defers low-AR-confidence positions to later steps, once the bidirectional decoder has resolved more surrounding context.
Through this deferral, DELTA-TTS commits even AR-hard positions at high confidence, producing the elevated distribution in Figure~\ref{fig:confidence_analysis}(b).

\subsection{Limitations and Future Work}\label{sec:appendix_limitations}

We view DELTA-TTS as a starting point for converting AR TTS models into dLLMs, rather than a finished system.
Specifically, the current setup determines the target length from a rule-based audio-to-character rate; this is a general challenge for dLLM-based generation, which maturing dLLM techniques may resolve.
The model also operates only in English; multilingual extension remains future work.
We also expect the conversion approach to transfer to other AR speech LMs and to richer multi-codebook tokenizers, and reward-based fine-tuning with metrics such as WER and speaker similarity is a natural next step.

\subsection{Broader Impacts}\label{sec:appendix_impacts}

This work is intended for academic research purposes. DELTA-TTS shows that TTS can be obtained from a pretrained AR backbone with $585$ hours of public data and $\sim$$15\%$ additional parameters, lowering the compute and data barrier for speech-synthesis research and bridging dLLM research between text and speech language modeling. As with any high-fidelity zero-shot TTS, the model carries misuse risks such as voice impersonation; we urge downstream users to obtain consent from any reference speaker and to clearly disclose synthetic audio.


\newpage

\end{document}